\documentstyle[12pt]{article}
\begin{document}
\def\kln{{\kappa_{L}}^{NC}}
\def\krn{{\kappa_{R}}^{NC}}
\def\klc{{\kappa_{L}}^{CC}}
\def\krc{{\kappa_{R}}^{CC}}
\def\ttz{{\mbox {$t$-${t}$-$Z$}\,}}
\def\bbz{{\mbox {$b$-${b}$-$Z$}\,}}
\def\tta{{\mbox {$t$-${t}$-$A$}\,}}
\def\bba{{\mbox {$b$-${b}$-$A$}\,}}
\def\tbw{{\mbox {$t$-${b}$-$W$}\,}}
\def\tltlz{{\mbox {$t_L$-$\overline{t_L}$-$Z$}\,}}
\def\blblz{{\mbox {$b_L$-$\overline{b_L}$-$Z$}\,}}
\def\brbrz{{\mbox {$b_R$-$\overline{b_R}$-$Z$}\,}}
\def\tlblw{{\mbox {$t_L$-$\overline{b_L}$-$W$}\,}}
\def\beq{\begin{equation}}
\def\enq{\end{equation}}
\def\ra{\rightarrow}
\def\ppbar{ {\rm p} \bar{{\rm p}} }
\def\ifb{ {\rm fb}^{-1} }
\def\ETslash{\not{\hbox{\kern-4pt $E_T$}}}
\def\ggtt{ q \bar q, \, g g \ra t \bar t }
\def\width{ \Gamma( t \ra b W^+) }
\def\doublespaced{\baselineskip=\normalbaselineskip\multiply
    \baselineskip by 150\divide\baselineskip by 100}
\doublespaced
\pagenumbering{arabic}
\begin{titlepage}
\begin{flushright}
\large{
MSUHEP-94/05\\
May, 1994}
\end{flushright}
\vspace{0.4cm}
\begin{center} \LARGE
{\bf PROBING THE COUPLINGS OF THE \\
TOP QUARK TO GAUGE BOSONS}
\end{center}
\begin{center}
{\bf Douglas O. Carlson, Ehab Malkawi and C.--P. Yuan}
\end{center}
\begin{center}
{
Department of Physics and Astronomy \\
Michigan State University \\
East Lansing, MI 48824}
\end{center}
\vspace{0.4cm}
\raggedbottom
\setcounter{page}{1}
\relax

\begin{abstract}
\noindent
We parameterize the non-universal couplings of \ttz and \tbw in
the electroweak chiral lagrangian approach, and examine the constraints
on these parameters from the LEP data. We also study how the SLC, Tevatron,
LHC and NLC can improve the measurement of these couplings. Different
symmetry breaking scenarios  imply different correlations among these
couplings, whose measurement will then provide a means
to probe the electroweak symmetry breaking sector.
\end{abstract}
\end{titlepage}
\newpage
\section{Introduction}
\indent

Studies on radiative corrections concluded that the mass ($m_t$)
of a Standard Model (SM) top quark has to be less than 200\,GeV \cite{holl}.
{}From the direct search at the Tevatron, assuming a SM top quark,
$m_t$ has to be larger than 131\,GeV \cite{D0}.
Recently, data were presented by the CDF group at the FNAL to
support the existence of a heavy top quark with mass
$m_{t}\sim 174\pm 20$\,GeV \cite{CDF}. However, there are no compelling
reasons to believe that the top quark couplings to other particles should be
of the SM nature. Because the top quark is heavy
relative to other observed fundamental particles, one expects that any
underlying theory at high energy scale $\Lambda \gg m_t$ will
easily reveal itself at low energy through the effective interactions
of the top quark to other particles.
Also, because the top quark mass is of the order of the Fermi scale
$v={(\sqrt{2}G_F)}^{-1/2}=246$\,GeV, which characterizes the
electroweak symmetry breaking scale, the
top quark would be useful in probing the symmetry breaking sector.

In this paper we constrain the effective couplings of the top quark to gauge
bosons using LEP data and discuss how the measurement of these couplings
can be improved by direct detection of the top quark
in either $e^-e^+$ ($e^- \gamma$) or hadron collisions.
In section 2 we examine what we have learned about the top
quark couplings from low energy data at LEP. In section 3 we study how to
probe the couplings $\klc$ and $\krc$ at the Tevatron and the LHC
from direct detection of the top quark.
In section 4 we discuss how the NLC can contribute to the measurement of
these couplings. Finally, in section 5 we discuss how to probe the
symmetry breaking sector by examining the correlations among the couplings
of the top quark to gauge bosons. Some conclusions
are also given in that section.
\section{Probing the Top Quark Couplings at LEP}
\indent

Since the top quark contributes to low energy data through radiative
corrections, one can indirectly probe the couplings of the top quark
to  gauge bosons at LEP. Taking the chiral lagrangian approach [4-13],
we systematically parameterize the interactions of the top quark
to gauge bosons at low energy using an effective lagrangian with the
non-linear realization of the symmetry ${SU(2)_{L}\times U(1)_Y} / {U(1)_{em}}$
\cite{ehab}. In the unitary gauge, it is
\begin{eqnarray}
{\cal L}= {\cal L}_{SM}&+&\frac{g}{4\cos \theta_W}\bar{t}\left
( \kln \gamma^{\mu}
(1-\gamma_{5})+\krn \gamma^{\mu}(1+\gamma_{5}) \right ) t\,Z_{\mu} \nonumber \\
 &+&\frac{g}{2\sqrt{2}}\bar{t}\left ( \klc\gamma^{\mu}(1-\gamma_{5})
+ \krc \gamma^{\mu}(1+\gamma_{5})\right ) b\,{W_{\mu}}^{+} \nonumber \\
&+&\frac{g}{2\sqrt{2}}\bar{b}\left ( {\klc}^{\dagger}\gamma^{\mu}
(1-\gamma_{5})+{\krc}^{\dagger}\gamma^{\mu}(1+\gamma_{5}) \right )
t\,{W_{\mu}}^{-} \label{eq3} \, ,
\end{eqnarray}
where ${\cal L}_{SM}$ is the SM lagrangian, $\kln$ and $\krn$ are two
arbitrary real parameters, $\klc$ and $\krc$ are two arbitrary complex
parameters and the superscript $NC$ and $CC$ denote neutral and charged
current respectively. In this work, we assume
the vertex \bbz is standard. The case where the vertex \bbz has a
non-standard effect comparable with the non-standard effect in the vertices
\ttz and \tbw, as expected in some Extended Technicolor models \cite {sekh},
will not be discussed here, but in a more detailed study in Ref.~\cite{ehab}.

The chiral lagrangian ${\cal L}$, as defined in Eq.~(\ref{eq3}),
has six independent parameters ({\mbox {$\kappa$'s}}) to be constrained
by low energy precision data. We will only consider the
insertion of {\mbox {$\kappa$'s}} once in one-loop diagrams by assuming
that these non-standard couplings are small, {\it i.e.},
$\kappa_{L,R}^{NC,CC} < {\cal O}(1)$. At one loop level the imaginary parts
of the couplings do not contribute in those LEP observables of interest.
Thus, hereafter we drop the imaginary part of these couplings,
which reduces the number of relevant parameters to four.
To the order ${\cal O}({m_t}^2\log {\Lambda}^2)$, $\krc$ does not contribute
to low energy data when ignoring the bottom quark mass, hence only the three
parameters $\kln$, $\krn$ and $\klc$ can be constrained by LEP data.

 A systematic approach can be implemented for such an
analysis based on the scheme used in Refs.~[16-18], where the
radiative corrections can be parameterized by 4 independent parameters,
three of those parameters ($\epsilon_1$, $\epsilon_2$, and $\epsilon_3$)
are proportional to the variables $S$, $U$ and $T$ \cite{pesk},
and the fourth  one ($\epsilon_b$) is due to the GIM violating
contribution in $Z\rightarrow b \overline{b}$ \cite{bar}.
These parameters are derived from four basic measured \mbox{observables},
$\Gamma_{\ell}$\,(the partial width of $Z$ to a charged lepton pair),
${A_{FB}}^{\ell}$\,(the forward-backward asymmetry at the $Z$ peak for
the charged lepton $\ell$), ${ M_{W} / M_{Z} }$ and $\Gamma_{b}$\,(the
partial width of $Z$ to a $b\overline{b}$ pair).

 Non-renormalizability of the effective lagrangian presents
a major issue of how to find a scheme to handle both the divergent and the
finite pieces in loop calculations \cite{burg,mart}. Such a problem arises
because one does not know the underlying theory, hence no matching can be
performed to extract the correct scheme to be used in the effective lagrangian
\cite{geor}. One approach is to associate the divergent piece in loop
calculations with a physical cut-off $\Lambda$, the upper scale at which the
effective lagrangian is valid \cite{pecc}. In the chiral lagrangian approach
this cut-off $\Lambda$ is taken to be $4\pi v \sim 3$\,TeV \cite{geor}. For
the finite piece no completely satisfactory approach is available \cite{burg}.

Performing the calculations in the unitary gauge, we calculate the
contribution to $\epsilon_1$ and $\epsilon_b$ due to the new interaction
terms in the chiral lagrangian (see Eq.~(\ref{eq3})) using the dimensional
regularization scheme and taking the bottom quark mass to be zero.
At the end of the calculation, we replace the divergent piece $1/\epsilon$ by
$\log({\Lambda^2}/{{m_t}^2})$ for $\epsilon = (4-n)/2$, where $n$ is the
space-time dimension. Since we are mainly interested in new physics
associated with the top quark couplings to gauge bosons, we shall restrict
ourselves to the {\it leading} contribution enhanced by the top quark
mass, {\it i.e.}, of the order of ${m_t}^{2}\log {\Lambda}^{2}$.

We find
\beq
\epsilon_1=\frac{G_F}{2\sqrt{2}{\pi}^2}3{m_t}^2
 (-\kln+\krn+\klc)\log{\frac{{\Lambda}^2}{{m_t}^2}}\,\, , \label{cal1}
\enq
\beq
\epsilon_b=\frac{G_F}{2\sqrt{2}{\pi}^2}{m_t}^2
\left ( -\frac{1}{4}\krn+\kln \right ) \log{\frac{{\Lambda}^2}{{m_t}^2}}
\,\, . \label{cal2}
\enq
Note that $\epsilon_2$ and $\epsilon_3$ do not contribute at this order.

To constrain these non-standard couplings we need to have both the
experimental values and the SM predictions of these {\mbox {$\epsilon$'s}},
which were obtained from Ref.~\cite{bar}.

Choosing $m_t=150$\,GeV, $m_H=100$\,GeV, and including both the SM and the
new physics contributions, we span the parameter space defined by
$-1 \leq \kln \leq 1 $, $-1 \leq \krn \leq 1 $ and $-1 \leq \klc \leq 1 $.
Within $95$\%~confidence level (C.L.), the allowed region of these three
parameters was found to form a thin slice in the specified volume.
The two-dimensional projections of this slice are shown in Figs.~(1)-(3).
These non-standard couplings {\mbox {($\kappa$'s)}} do exhibit some
interesting features:
\begin{itemize}
\item [1)]
As a function of the top quark mass,
the allowed volume for the top quark
couplings to gauge bosons shrinks as the top quark becomes
more massive.
\item [2)]
 New physics prefers positive $\kln$,
see Figs.~(1) and (2).
$\kln$ is constrained to be
within $-0.3$ to 0.6 ($-0.2$ to 0.5) for
a 150 (175)\,GeV top quark.
\item [3)]
New physics prefers  $\klc \approx -\krn$.
This is clearly shown in Fig.~(3) which is
the projection of the allowed volume
in the $\klc$ and $\krn$ plane.
\end{itemize}

In Ref.~\cite{pczh}, a similar analysis has been
carried out by Peccei {\it et al}. However, in their analysis they did
not include the charged current contribution and assumed only the vertex
\ttz gives large non-standard effects. The allowed region they found  simply
corresponds, in our analysis, to the region defined by the intersection
of the allowed volume  and the plane $\klc = 0$. This gives a small area
confined in the vicinity of the line $\kln = \krn$. (This is obtained by
setting $\klc=0$ in Eq.~(\ref{cal1}).) In this case we note that the length of
the allowed area is merely determined by $\epsilon_b$.

To conclude, assuming \bbz vertex is not modified, we found that $\kln$ is
already constrained at the  95\% C.L. to be
$-0.3 < \kln < 0.6$ ($-0.2 < \kln < 0.5$) by LEP data
for a 150 (175)\,GeV top quark. Although $\krn$ and $\klc$ are allowed to be
in the full range of $\pm 1.0$, the precision LEP data do impose some
correlations among $\kln$, $\krn$ and $\klc$. Note that $\krc$ does not
contribute to the LEP observables of interest in the limit of $m_b=0$.

At the SLC, with expected better measurement of the
left-right cross section asymmetry $A_{LR}$ in $Z$ production
with a longitudinally polarized electron beam,
one can further constrain these $\kappa$'s  \cite{ehab}.
\section{At the Tevatron and the LHC }
\indent

In this section, we study how to constrain the non-standard couplings
of the top quark to gauge bosons
from direct detection of the top quark at hadron colliders.

At the Tevatron and the LHC, heavy top quarks are
predominantly produced from the QCD process
$gg, q \bar q \ra t \bar t$ and the $W$-gluon fusion process
$qg (Wg) \ra t \bar{b}, \bar{t} b$. In the former process, one can probe
$\klc$ and $\krc$ from the decay of the top quark to a bottom quark and
a $W$ boson. In the latter process, these non-standard couplings can be
measured by simply counting the production rates of signal events with a
single $t$ or $\bar t$. More details can be found in Ref.~\cite{anlrev}.

To probe $\klc$ and $\krc$ from the decay of the
top quark to a bottom quark and a $W$ boson, one needs to measure the
polarization of the $W$ boson. For a massless $b$, the $W$ boson from top
quark decay can only be either longitudinally or left-handed polarized for
a left-handed charged current ($\krc=0$). For a right-handed
charged current ($\klc=-1$) the $W$ boson can only be either longitudinally
or right-handed polarized. (Note that the handedness of the $W$ boson is
reversed for a massless $\bar b$ from $\bar t$ decays.)
In all cases the fraction of longitudinal $W$
from top quark decay is enhanced by ${m_t}^2/{2{M_W}^2}$ as compared
to the fraction of transversely polarized $W$. Therefore, for a more massive
top quark, it is more difficult to untangle the $\klc$ and $\krc$
contributions. The $W$ polarization measurement can be done by measuring the
invariant mass ($m_{b\ell}$) of the bottom quark
and the charged lepton from the decay of top quark \cite{kanetop}.
We note that this method does not require knowing the longitudinal momentum
(with two-fold ambiguity) of the neutrino from $W$ decay to reconstruct
the rest frame of the $W$ boson in the rest frame of the top quark.

Consider the (upgraded) Tevatron as a $\ppbar$ collider at $\sqrt{S}= 2$
or 3.5 TeV, with an integrated luminosity of 1 or 10\,$\rm{fb}^{-1}$.
Unless specified otherwise, we will give event numbers for a 175\,GeV top
quark and an integrated luminosity of 1\,$\rm{fb}^{-1}$.

The cross section of the QCD process $gg,q\bar{q}\ra t \bar t$
is about 7 (29)\,pb at a $\sqrt{S}= 2$ (3.5)\,TeV collider.
In order to measure $\klc$ and $\krc$ we have to study the decay kinematics
of the reconstructed $t$ and/or $\bar t$.
For simplicity, let's consider the $\ell^\pm \, + \geq 3\, {\rm jet}$
decay mode, whose branching ratio is
$Br=2 {\frac{2}{9}} {\frac{6}{9}} = \frac{8}{27}$, for
$\ell^+=e^+ \,{\rm or}\,\mu^+$. We assume an experimental detection
efficiency, which includes both the kinematic acceptance and the efficiency
of $b$-tagging, of 15\% for the $t \bar t$ event. We further assume that
there is no ambiguity in picking up the right $b$ ($\bar b$)
to combine with the charged lepton $\ell^+$ ($\ell^-$)
to reconstruct $t$ ($\bar t$). In total, there are $7\,{\rm pb}\,\times
\, 10^3\,{\rm pb}^{-1}\, \times \,{\frac{8}{27}}\,\times \,0.15=300$
reconstructed $t \bar t$ events to be used in measuring
$\klc$ and $\krc$ at $\sqrt{S}= 2$\,TeV. The same calculation at
$\sqrt{S}= 3.5$\,TeV yields $1300$ reconstructed $t \bar t$ events.
Given the number of reconstructed top quark events,
one can in principle fit the $m_{b\ell}$ distribution to measure
$\klc$ and $\krc$. We note that the polarization of the $W$ boson can also be
studied from the distribution of $\cos \theta^*_\ell$,
where $\theta^*_\ell$ is the polar angle of $\ell$ in the rest frame of the
$W$ boson whose z-axis is the $W$ bosons moving direction in the rest
frame of the top quark \cite{kanetop}.
For a massless $b$, $\cos \theta^*_\ell$ is related to ${m_{b\ell}}^2$ by
\beq
{\cos \theta^*_{\ell}}
\simeq {{{2 {m_{b\ell}}^2}\over{{m_t}^2-{M_W}^2}} - 1}\, .
\enq

However, in reality, the momenta of the bottom quark and the
charged lepton will be smeared by
the detector effects and the most serious problem in this analysis is
the identification of the right $b$ to reconstruct $t$.
There are two strategies to improve the efficiency of identifying
the right $b$.  One is to demand a large invariant mass of the $t \bar t$
system so that $t$ is boosted and its decay products are collimated.
Namely, the right $b$ will be moving closer to the lepton from $t$ decay.
This can be easily enforced by demanding lepton $\ell$ with large transverse
momentum. Another is to identify the non-isolated lepton from $\bar b$ decay
(with a branching ratio $Br(\bar b \ra \mu^{+} X) \sim 10\%$). Both
of these methods will further reduce the reconstructed signal rate by
an order of magnitude. How will these affect our conclusion on
the determination of the non-universal couplings $\klc$ and $\krc$?
This cannot be answered in the absence of detailed Monte Carlo studies.

Here we propose to probe the couplings $\klc$ and $\krc$ by measuring the
production rate of the single-top quark events. A single-top quark event
 can be produced from either the $W$-gluon fusion process
$q g \, (W^+g) \ra t \bar{b} X$, or the Drell-Yan type process
$q \bar q \ra W^* \ra t \bar b$.
Including both the single-$t$ and single-${\bar t}$ events,
for a 2 (3.5)\,TeV collider, the $W$-gluon fusion rate is 2 (16)\,pb;
the Drell-Yan type rate is 0.6 (1.5)\,pb.
The Drell-Yan type event is easily separated from the
$W$-gluon fusion event, therefore will not be considered
hereafter \cite{carlson}. For the decay mode of
$t \ra b W^+ \ra b \ell^+ \mu$, with $\ell^+=e^+ \,{\rm or}\,\nu^+$,
the branching ratio of interest is $Br=\frac{2}{9}$.
The kinematic acceptance of this event at $\sqrt{S}= 2$\,TeV is
found to be $0.55$ \cite{carlson}. If the efficiency of $b$-tagging is 30\%,
there will be $2\,{\rm pb}\,\times \,10^3\,{\rm pb}^{-1}\,
\times \,{\frac{2}{9}}\,\times \,0.55\, \times \,0.3=75$ single-top quark
events reconstructed.   At $\sqrt{S}= 3.5$\,TeV the kinematic acceptance
of this event is $0.50$ which, from the above calculation yields
about $530$ reconstructed events. Based on statistical
error alone, this corresponds to a 12\% and 4\% measurement on the single-top
cross section. A factor of 10 increase in the luminosity of
the collider can improve the measurement by a factor of 3 statistically.

Taking into account the theoretical uncertainties, we examine two scenarios:
20\% and 50\% error on the measurement of the single-top cross section,
which depends on both $\klc$ and $\krc$. (Here we assume the experimental data
agrees with the SM prediction within 20\% (50\%).) We found that for a
$175\,$GeV top quark $\klc$ and $\krc$ are well constrained inside
the region bounded by two (approximate) ellipses, as shown in Fig.~(4).
These results are not sensitive to
the energies of the colliders considered here.

The top quark produced from the $W$-gluon fusion process
is almost one hundred percent left-handed (right-handed) polarized
for a left-handed (right-handed) $\tbw$ vertex, therefore
the charged lepton $\ell^+$ from $t$ decay has a harder momentum
in a right-handed $\tbw$ coupling than in a left-handed coupling.
(Note that the couplings of light-fermions to $W$ boson have been well tested
from the low energy data to be left-handed as described in the SM.)
This difference becomes smaller when the top quark is more massive  because
the $W$ boson from the top quark decay tends to be more
longitudinally polarized.

A right-handed charged current is absent in a linearly $SU(2)_L$ invariant
gauge theory with massless bottom quark. In this case,  $\krc=0$,
then $\klc$ can be constrained to within about
$-0.08 < \klc < 0.03$ ($-0.20 < \klc < 0.08$) with a 20\% (50\%) error on
the measurement of the the single-top quark production rate at the Tevatron.
This means that if we interpret {\mbox {($1+\klc$)}} as the CKM matrix
element $V_{tb}$, then $V_{tb}$ can be bounded as $V_{tb} > 0.9$ (or 0.8)
for a 20\% (or 50\%)  error on the measurement of the single-top production
rate. Recall that if there are more than three generations, within  90\% C.L.,
$V_{tb}$ can be anywhere between 0 and 0.9995 from low energy data
\cite{book}. This measurement can therefore provide useful information on
 possible additional fermion generations.

We expect the LHC can provide similar or better bounds on these
non-standard couplings when detail analyses are available.

\section{ At the NLC }
\indent

The best place to probe $\kln$ and $\krn$ associated with the
\ttz coupling is at the NLC through $e^- e^+ \ra A, Z \ra t \bar{t}$.
(We use NLC to represent a generic $e^-e^+$ supercollider \cite{nlc}.)
A detail Monte Carlo study on the measurement of these couplings at the NLC
including detector effects and initial state radiation
can be found in Ref.~\cite{gal}. The bounds were obtained by studying the
angular distribution and the polarization of the top quark
produced in $e^- e^+$ collisions. Assuming a 50 $\rm{fb}^{-1}$
luminosity at $\sqrt{S}=500$\,GeV, we concluded that
within 90\% confidence level, it should be possible to measure
$\kln$ to within about 8\%, while $\krn$ can be known to within about 18\%.
A $1\,$TeV machine can do better than a $500\,$GeV machine in
determining $\kln$ and $\krn$ because the relative sizes of the
$t_R {(\overline{t})}_R$  and $t_L {(\overline{t})}_L$
production rates become small and the polarization of the $t \bar t$ pair
is purer. Namely, it's more likely to produce either
a $t_L {(\overline{t})}_R$ or a $t_R {(\overline{t})}_L$ pair.
A purer polarization of the $t \bar t$ pair makes $\kln$ and $\krn$
better determined. (The purity of the $t \bar t$ polarization
can be further improved by polarizing the electron beam.)
Furthermore, the top quark is
boosted more in a $1\,$TeV machine thereby allowing a better
determination of its polar angle in the $t \bar t$ system
because it is easier to find the right $b$ associated with the lepton to
reconstruct the top quark moving direction.

Finally, we remark that at the NLC, $\klc$ and $\krc$ can be studied
either from the decay of the top quark pair or from the single-top
quark production process; W-photon fusion
process $e^{-}e^{+}(W\gamma) \ra t X $,
or $ e^{-}\gamma (W\gamma) \ra {\bar t} X$, which is similar to the
$W$-gluon fusion process in hadron collisions.

\section{Discussion and Conclusions}
\indent

Different symmetry breaking scenarios will imply different correlations
among the couplings of the top quark
to gauge bosons. By examining these
correlations one may be able probe the electroweak
symmetry breaking sector.
To illustrate how a specific symmetry breaking mechanism might affect
these couplings in the effective lagrangian,
we consider the SM with a heavy Higgs boson as the full
theory, and perform matching between the underlying theory (SM)
and the effective lagrangian. We find \cite{ehab}
\beq
{\kappa_{L}}^{NC}=2{\kappa_{L}}^{CC}= \frac {G_F}{2\sqrt{2}{\pi}^2}
 \left(\frac{-1}{8}\right){m_t}^2
 \log\left(\frac{{m_H}^{2}}{{m_t}^{2}}\right)\, ,
\enq
\beq
{\kappa_{R}}^{NC}=\frac{G_F}{2\sqrt{2}{\pi}^2}
\frac{1}{8}{m_t}^2
 \log\left(\frac{{m_H}^{2}}{{m_t}^{2}}\right)\, ,
\enq
\beq
{\kappa_{R}}^{CC}=0\, ,
\enq
at the scale $m_t$. Note that due to the Ward identities
associated with the photon field there can be no non-universal contribution
to either the \bba or \tta vertex after renormalizing the fine structure
constant $\alpha$. This can be explicitly checked in this model.
Furthermore, up to the order of ${m_t}^{2}\log{m_H}^{2}$, the
vertex \bbz is not modified.

{}From this example we learn that the effective non-standard couplings
of the top quark to gauge bosons arising from a heavy Higgs boson are
correlated in a specific way, namely
\beq
\kln=2\klc=-\krn \,\, \, {\rm{and}}\,\, \krc=0 \, .
\enq
Hence, if the couplings of a heavy top quark to gauge bosons are measured
and exhibit large deviations from these relations, then it is likely that
the electroweak symmetry breaking is not due to the standard  Higgs mechanism
with a heavy SM Higgs boson. This illustrates how one may be able
to probe the symmetry breaking sector by measuring the effective couplings
of the top quark to gauge bosons.

In conclusion, assuming the \bbz vertex is not modified, from LEP data,
we found that $\kln$ is already constrained to be
$ -0.3 < \kln <0.6\, ( -0.2 <\kln < 0.5 )$ at the 95\% C.L. for a
150 (175)\,GeV top quark. Despite the fact that $\krn$, $\klc$ and $\krc$
are not constrained in the range of $\pm 1.0$, LEP data does impose some
correlations among the couplings $\kln$, $\krn$ and $\klc$.
But, $\krc$ is not bounded because it
does not contribute to the LEP observables at the order
of ${m_t}^2\log {\Lambda}^2$.

Undoubtedly, direct detection of the top quark at the Tevatron,
the LHC and the NLC is crucial to measuring the couplings of
\tbw and \ttz. At hadron colliders, $\klc$ and $\krc$ can be measured by
studying the polarization of the $W$ boson from top quark decay.
They can also be measured simply from the production
rate of the single-top quark event.
The NLC is the best machine to measure $\kln$ and $\krn$ which can be
measured from studying the angular distribution and the polarization
of the top quark produced in $e^- e^+$ collisions.

\section*{ Acknowledgments }
\indent

We thank
R. Brock, R.S. Chivukula, M. Einhorn, K. Lane, E. Nardi,
E.H. Simmons, M. Wise and Y.--P. Yao
for helpful discussions.
This work was supported in part by an NSF grant No. PHY-9309902.

\newpage
\section*{Figure Captions}
\noindent
Fig. 1.\\
Two-dimensional projection in the plane of $\kln$ and $\krn$, for
$m_t=150$\,GeV, $m_H=100$\,GeV.
\vspace{0.3cm}

\noindent
Fig. 2.\\
Two-dimensional projection in the plane of $\kln$ and $\klc$, for
$m_t=150$\,GeV, $m_H=100$\,GeV.
\vspace{0.3cm}

\noindent
Fig. 3.\\
Two-dimensional projection in the plane of $\krn$ and $\klc$, for
$m_t=150$\,GeV, $m_H=100$\,GeV.
\vspace{0.3cm}

\noindent
Fig. 4.\\
The allowed
$|\krc|$ and $\klc$ are bounded within
the two dashed (solid) lines
for a 20\% (50\%)  error on the measurement
of the single-top production rate, for a 175 GeV top quark.
\end{document}